# Spatially and Temporally Resolved Mapping of Contact Electrification on Stand-Alone Ultrathin Glass Materials via Kelvin Probe Force Microscopy

*Aayush Nayyar[1], Ruizhe Yang[1,2 †], Vashin Gautham[1], Sagnik Das[3], Haiqing Lin[3], Andrew C. Antony[4], Dean Thelen[4], Mayukh Nath[4], Gabriel Agnello[4\*],  Jun Liu[1\*]*

[1]Department of Mechanical and Aerospace Engineering and RENEW (Research and Education in Energy, Environment and Water) Institute, University at Buffalo, The State University of New York, Buffalo, New York 14260, United States

[2]Current address: Pritzker School of Molecular Engineering, The University of Chicago, Chicago, Illinois 60637, United States

[3]Department of Chemical and Biological Engineering, University at Buffalo, The State University of New York, Buffalo, New York 14260, United States

[4]Science & Technology Division, Corning Incorporated, Corning, United States

**ABSTRACT**

Contact electrification (CE) remains a critical challenge in advanced material technologies where uncontrolled surface charging can compromise manufacturability, reliability and performance of the materials for practical applications. Ultrathin glass with micrometer-scale thickness is a state-of-the-art specialty oxide material for flexible touchscreen in new-generation electronic devices. Despite extensive studies on CE on thermally grown oxide thin films, the physical and chemical properties of the stand-alone ultrathin oxide materials, could be very different and thus lead to distinct CE behaviors. Such behaviors have not been experimentally investigated due to the challenge of their ultrathin form factor as well as the lack of experimental methods that would allow the successful study of CE on stand-alone ultrathin glass materials. Here, we, for the first time, visualize and quantify CE-induced surface charges on ultrathin glasses using sideband-mode Kelvin probe force microscopy (KPFM). To enable the KPFM measurement, we have established experimental strategies including electrode preparation enabling the measuring circuit, and surface cleaning procedures improving surface activation and hydrophilicity. Nano-sized atomic force microscopy (AFM) probes were used to scan and induced triboelectric charges on the stand-alone glass surfaces with a variety of thicknesses (30-100 μm) under ultra-pure $N_2$ conditions. Time-dependent measurements reveal the surface charges on 30 μm-thick glass sample decay from 4.47 V to 0.37 V in 240 minutes. Moreover, it is found that electrostatic charges exhibit a capacitor-like discharging behavior primarily through the bulk material yielding a long relaxation time constant of ~41 minutes, which is different from the lateral surface discharging behaviors in thermally grown $SiO_2$ thin film reported previously. Furthermore, the thickness-dependent surface charging effect was characterized for the ultrathin glass substrates, where the change in contact potential difference between the charged and uncharged region ($\Delta V_{CPD}$) was found to remain nearly constant across this thickness range from 1.39 ± 0.17 V at 30 μm to 1.34 ± 0.29 V at 100 μm. A self-capacitance analytical model was developed and employed to estimate the corresponding surface charge density ($\sigma$), yielding comparable values of 136.26 ± 16.25 μC/m$^2$ at 30 μm and 131.44 ± 28.41 μC/m$^2$ at 100 μm. Additionally, the external bias applied to the AFM tips can be used to enhance, suppress, or invert the intrinsic CE response of glass materials. This work extends nanoscale CE characterization beyond oxide thin films to stand-alone oxide materials, providing a framework to understand and manipulate electrostatic charging in glass systems for practical applications.

**INTRODUCTION**

Over the centuries, advancements in material processing and manufacturing techniques have led to the widespread use of glass materials in human history. Today, silicate glasses are incorporated in applications such as optical communications,[1, 2] display panels,[3, 4] touchscreen in mobile electronics,[5, 6] and energy harvesting devices.[7, 8] Despite its extensive commercial applications and the level of engineering sophistication involved in its production, there remain several unanswered fundamental questions regarding the physical and chemical behavior of glass. One such phenomenon that is continuously under investigation is contact electrification (CE) (or triboelectrification), the process by which charge transfers between two surfaces via material contact and separation.[9-11]

Uncontrolled surface charging of glass can lead to dust attraction, organic particle contamination, decreased adhesion between film and the glass substrate, and structural failures.[12, 13] In the flat panel glass display industry for example, significant surface charge accumulation can occur due to the unavoidable "material A"-glass contact that occurs during the handling and transportation of the panels via conveyors, rollers, and robotic grippers.[14, 15] These charges can be highly localized and increase the risk of dielectric breakdown in glass panels which can compromise structural integrity, damage underlying components, and cause electrostatic discharge (ESD) leading to short circuits and point defects.[16-18] Such defects negatively impact the performance and reliability of the manufactured products and result in substantial material losses, making electrostatic control a critical challenge in large-scale glass manufacturing.[19] Thus, a fundamental understanding of the mechanism by which contact electrification occurs in glass, along with the factors that affect it, is essential to develop methods that can mitigate surface charging and improve the efficiency of manufacturing processes as well as the reliability of glass-based technologies.

Despite being well documented and applied in several technologies, the microscopic origin, spatial distribution, and relaxation of charges due to CE remain debated.[20-22] Over time, three main theories explaining the CE mechanism have been developed: electronic transfer, ionic transfer, and mass transfer.[23-25] Electron transfer is widely accepted as the main mechanism for CE in metal-metal and metal-semiconductor based systems, where charging occurs due to the transfer of electrons based on relative work functions or Fermi levels.[26, 27] For insulators such as glass, however, both electron and ion transfer theories present strong arguments. While the existence of electronic surface states at energy levels between the valence band maximum and conduction band minimum may contribute to electron transfer mechanisms, these states can be passivated by ambient $H_2O$ molecules which react with the glass surface to form silanol groups (-SiOH).[28] In addition to hydroxyl ions at the surfaces, other elements from multicomponent glass such as network modifying $Na^+$ or $K^+$ species can act as mobile ions which contribute to the ionic transfer mechanism, thus presenting a strong argument for an ion transfer based charging mechanism.[21, 29, 30]

The problem becomes even more challenging for glass, as its bulk-composition and surface properties can vary significantly depending on the application. Numerous experimental, computational, and simulation studies have been conducted to characterize CE on glass surfaces. Macroscopic techniques involving metallic or insulating rollers[15, 16] and rolling sphere tests (RST)[19, 31] have been used to induce and measure the charges on glass substrates. The effect of humidity, structural defects due to varying glass compositions, and surface treatment processes on CE have also been studied extensively. Relative humidity strongly influences surface resistivity and charge dissipation by controlling the adsorption, structure, and hydrogen-bonding interactions of interfacial water layers on the glass surface.[32] Treating the glass surface with varying pH solutions can also change the topmost surface and the structure of the water layer, affecting the charge-discharge behavior.[33]

Density functional theory (DFT) and molecular dynamics (MD) based computational techniques have also been used to study the origin and stability of surfaces charges. For example, varying the contacting metal (Pt, Au, or Al) in a $SiO_2$/metal DFT calculation changes the equilibrium electronic charge transfer between the surfaces.[34] MD simulations have further shown that the electrostatic charging behavior of Ca-Al-Si (CAS) glasses is closely related to their surface atomic structure and spatial distribution of ionic species.[35] While these studies present invaluable insights into the atomic and electronic details of CE, high-resolution experimental tools are needed to empirically visualize how CE induced charges are generated, distributed, and dissipated on insulating substrates.

Kelvin Probe Force Microscopy (KPFM) has emerged as a powerful tool for investigating CE in thin dielectric films. With a combination of contact-mode based AFM and KPFM, contact charging can be induced, controlled, and characterized systematically.[36, 37] However, existing studies utilizing this KPFM technique have been largely restricted to thermally grown silicon oxide layers with thicknesses of a few hundred nanometers on conductive silicon substrates. Such systems differ substantially from the stand-alone glasses used in commercial technologies which typically range from tens of micrometers to millimeters in thickness. Furthermore, most commercially available glass panels, based on specific applications, can exhibit a wide variety of compositions such as soda-lime, calcium-aluminosilicates, and alkali containing borosilicates. Applying the KPFM technique in such systems poses additional challenges. Thick insulating substrates reduce the tip-sample capacitance, making surface potential measurements less reliable. Moreover, uncontrolled surface contamination on stand-alone glass systems can significantly affect the magnitude, distribution, and decay of the charges induced due to CE, thus hindering accuracy and reproducibility.

In this work, we address these challenges by employing KPFM to investigate CE in ultrathin (30-100 μm) glasses. By systematically investigating the influence of glass thickness and surface treatment on nanoscale charge behavior, our study extends KPFM based CE analysis beyond thin oxide films, providing new insights into the charging and discharging dynamics of silicate glass.

**EXPERIMENTAL SECTION**

**Glass samples**

A commercially available magnesium alumino-sodium silicate glass (Corning Gorilla Glass 2™) was chosen for this study. Glasses were melted and formed to approximately 300 μm, then thinned to target thickness using standard HF based etchants. Chemical strengthening processes were performed on all samples using a standard ion exchange (IOX) process.[38, 39] These processes resulted in chemical alteration layers at the surfaces of the samples approximately ~15-20% of the total thickness of the glass.

**Gold (Au) Thin Film Deposition**

An E-beam evaporator with glancing angle deposition (GLAD) with multi-position sample mount arm from Kurt J. Lesker Company AXXIS was used for depositing thin films. A 5 nm seed layer of chromium and 50 nm layer of gold was deposited at the bottom of the glass surface. The system was first pumped down to a pressure of <$10^{-6}$ Torr. The top rotating plate was spun at a rate of 20 rpm, and the target deposition rate was set as 0.3 Ås$^{-1}$. Two quartz thickness monitors were used to detect the thickness of the thin film during deposition.

**Sample Preparation**

The glass samples were first cleaned with a standard wash by rinsing them with Acetone, Methanol, Isopropyl Alcohol, and ultra-pure DI water (18 mOhm.cm resistivity), and were then immersed in a 1M KOH solution at 80º C for 20 minutes. After KOH treatment, samples were transferred into a clean beaker with ultra-pure DI water and were ultrasonicated at 50º C for 10 minutes. Samples were then removed and dried with a jet of ultrapure nitrogen gas (Airgas research grade) and immediately transferred to a plastic box for testing.

**Atomic Force Microscopy Characterization**

Atomic force microscopy (AFM) characterization was conducted using a Park FX40 AFM system from Park Systems. PPP-NCHR (k=~42 N/m, f=~330 kHz), a n-Si based AFM probe with a tip radius of <10 nm from Nanosensors was used for contact charging and subsequent KPFM analysis. The AFM system was placed inside a controlled $N_2$ purged environment (Airgas research grade) and low humidity (<0.1 ppm $H_2O$). The AFM was first operated in contact mode over a 2×0.5 μm$^2$ area on the glass surface at a 50 nN contact force and a scan rate of 0.4 Hz to induce surface charges through contact electrification. KPFM scan was conducted over a larger area of 4×1 μm$^2$ to map the surface potential and topography at the surface of the glass samples.

**Contact Angle Measurement**

A ramé-hart Contact Angle Goniometer (Model 190) was used to measure the contact angle of DI water droplets on DI sonicated, standard washed, and KOH treated 30 μm glass samples.

## RESULTS AND DISCUSSION

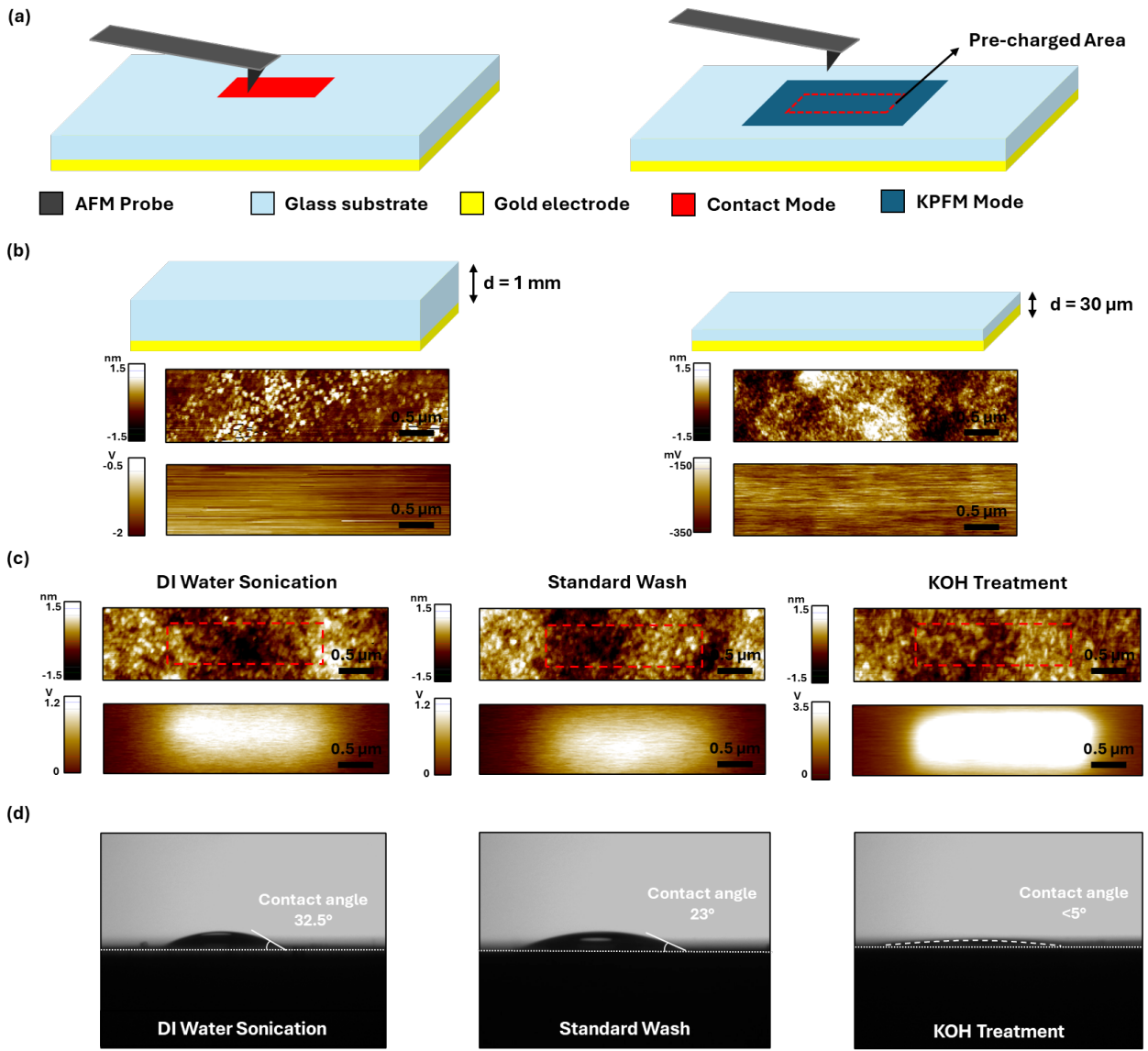

**Figure 1.** Experimental setup and effects of surface treatments on surface charging and KPFM response in ultrathin glass. (a) Schematic of AFM contact charging and subsequent KPFM measurement region. (b) Topography and surface potential images for 1 mm thick glass slide (Left) and 30 μm stand-alone glass (Right). (c) Topography and surface potential images for 30 μm glass with DI water sonication (Left), standard wash (Middle), and KOH treatment (Right). (d) Contact angle measurement demonstrating increasing hydrophilicity with different treatment conditions.

Figure 1a illustrates the experimental design for characterization of CE induced surface charges. First, a 50 nm gold thin film was deposited onto all glass substrates using an e-beam evaporator to form the bottom electrode for the KPFM scan, forming a parallel plate capacitor with the AFM tip.[40] The back electrode provides a well-defined electrical ground which is critical in avoiding residual charges within the bulk that can distort the measured surface potential, especially in thick dielectric substrates where the tip-sample capacitance is already weak.

Sideband-mode KPFM, a frequency-modulated KPFM technique, enables the mapping of surface potential and work function by detecting short-range force gradient interactions at the AFM probe tip. On applying an AC voltage at $\omega_{AC}$ to a cantilever mechanically oscillating at $\omega_o$, sidebands are generated in the frequency spectrum at $\omega_0 \pm \omega_{AC}$. The electrostatic force ($F_{es}$) between the tip and the sample can be expressed as: [41]

$$F_{es}(z,t) = -\frac{1}{2}\frac{\partial C(z)}{\partial z}[(V_{DC} \pm V_{CPD}) + V_{AC}\sin(\omega_{AC}t)]^2 \quad (1)$$

Where $z$ is the direction normal to sample surface, $t$ is time, $\partial C / \partial z$ is the capacitance gradient, $V_{DC}$ is the applied DC bias, $V_{CPD}$ is the contact potential difference, and $V_{AC}$ is the applied AC bias. This sideband response can be nullified by applying the DC voltage bias ($V_{DC}$) to compensate for the contact potential difference ($V_{CPD}$) between the tip and the sample, allowing for the direct quantification of the surface potential.

Compared to amplitude modulated (AM) KPFM, where long-range forces interactions such as that between the cantilever and the sample surface, sideband-mode KPFM is only affected by short-range electrostatic force interactions which is highly localized at the tip apex, thus offering higher spatial resolution and reduced topographic cross-talk[42].

To induce surface charges through contact electrification, a n-Si tip-based AFM probe was first scanned over a 2×0.5 µm² area on the glass surface at a 50 nN contact force and a scan rate of 0.4 Hz. The surface topography, charge distribution, and magnitude were then measured by conducting a KPFM scan over an enlarged area of 4×1 µm². The use of contact charging in a small region followed by a larger area scan enables the direct visualization of localized surface charges, lateral spreading, and decay dynamics over time.

Atmospheric moisture deposits a thin film of water on the sample surface can affect surface conductivity and CE.[43, 44] Thus, to reduce charge dissipation due to moisture and improving the accuracy and stability of the experiments, the KPFM measurements were conducted in a nitrogen purged environment with <0.1 ppm $O_2$ and $H_2O$.

A glass slide of 1mm thickness and commercially available glass panel of 30 µm thickness were used to test the effect of reducing glass thickness on KPFM signal intensity (Figure 1b). Due to strong attenuation of tip-sample capacitance by the insulating bulk, the measured surface potential signal for the 1 mm glass slide exhibits random values with a line averaged surface potential of -1.57 ± 0.15 V. For the 30 µm free-standing glass however, an enhancement in the signal was detected, with a surface potential of -0.25 ± 0.016 V along with a higher signal-to-noise ratio.

To ensure surface cleanliness, reproducibility, and electrostatic stability during measurement, a standard operating procedure was developed. The glass samples were first cleaned with a standard wash by rinsing them with Acetone, Methanol, Isopropyl Alcohol, and ultra-pure DI water (18 mOhm.cm resistivity) to remove organic contaminants from the sample surface. The samples were then immersed in a 1M KOH solution at 80° C for 20 minutes.

The KOH treatment acts as a chemical etching and surface activation step that removes deeper organic residues and ionic contaminants and activates the surface for better charge trapping and tip-sample interaction during CE. The thickness of glass removed during the process is estimated to be in the order of a few nanometers. After the KOH treatment, samples were ultrasonicated with ultra-pure DI water at 50° C for 10 minutes to remove residual ions. Finally, the samples were dried with a jet of ultrapure nitrogen gas (Airgas research grade) and transferred onto the sample stage for measurement. As shown in Figure 1c, by following the surface treatment procedure, the root mean square (RMS) roughness decreased from 0.5 nm for DI water sonicated and standard washed samples, to 0.3 nm with the KOH treatment. A cleaner surface is thus obtained which is essential for contact charging and KPFM measurements. An enhancement in surface charging is also observed, with the change in contact potential difference $\Delta V_{CPD}$ increasing from 0.98 V and 1.44 V for DI water sonicated and standard washed sample respectively, to 2.65 V for the KOH treated sample. Figure 1d shows the systematic reduction in contact angle of a water droplet at the sample surface from 32.5° after DI water sonication, to 23° after standard wash, to <5° after KOH treatment. It is evident that the glass surface becomes more hydrophilic after each step.

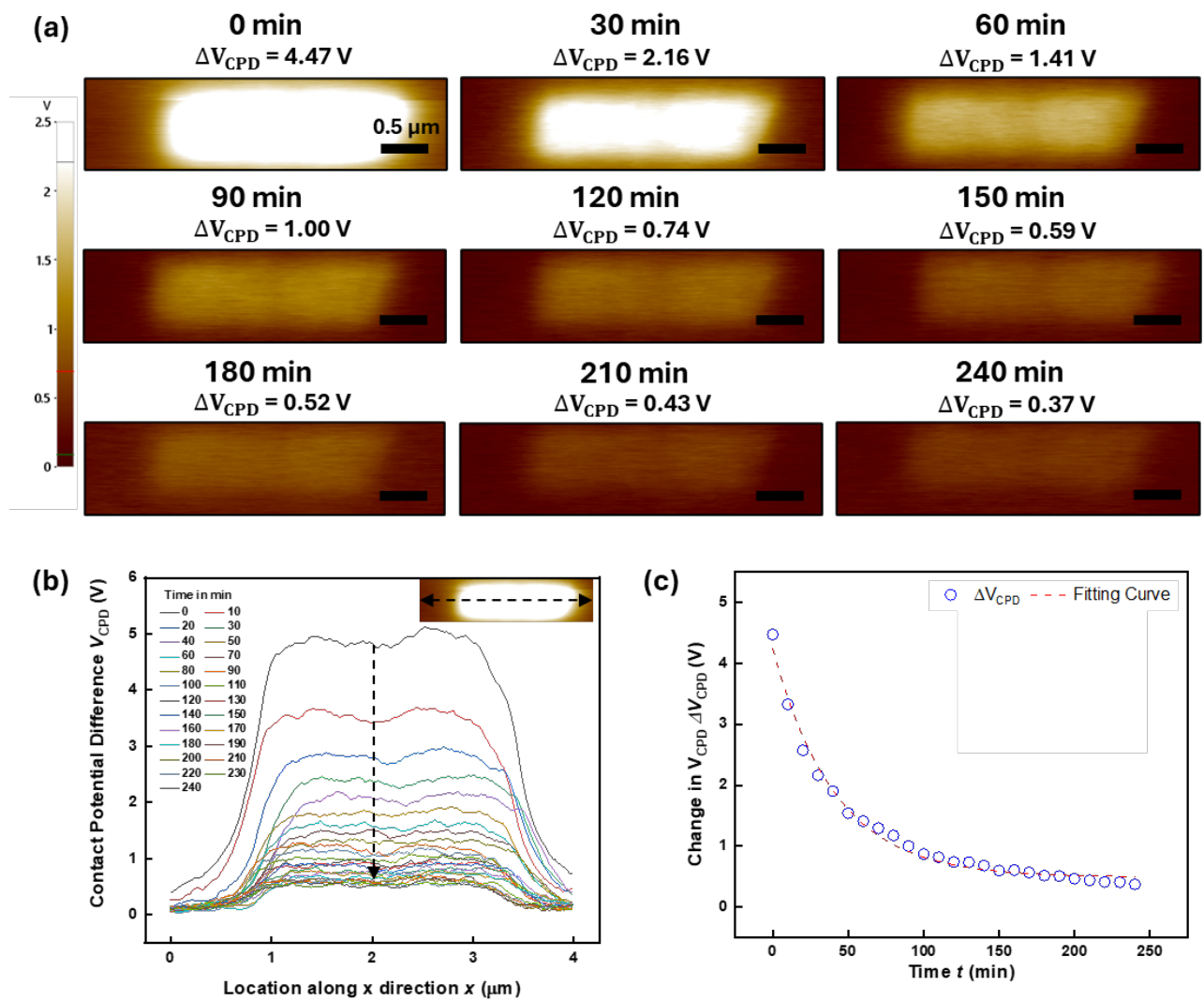

**Figure 2.** Time-dependent decay of CE-induced charges on 30 μm glass. (a) Time-lapse KPFM surface potential images of the same area recorded over 4 hours after three contact charging cycles. (b) Cross-sectional $V_{CPD}$ line profiles along central line scan (inset) at intervals of 10 minutes. (c) Time evolution of $\Delta V_{CPD}$ calculated from histogram peak difference (see Fig. S1 in Supporting Information for more details), fitted with an exponential decay model (Eq. 2).

The decay behavior of the charges after three contact scan cycles on a 30 μm thick glass sample was monitored at 10-minute intervals over a period of 4 hours as shown in Figure 2. At each interval, $\Delta V_{CPD}$ was calculated by comparing the peaks of the charged and uncharged domains in the histogram distribution of the measured surface potential (Figure S 1-2). After four hours, the measured $\Delta V_{CPD}$ is found to decreased from 4.47 V to 0.37 V. The $V_{CPD}$ line profile along the central line scan at each interval is shown in Figure 2b. Previous studies have reported a progressive broadening of the charged area on a 200 nm thick thermally grown $SiO_2$ layer on an n-Si wafer due to dominant surface diffusion. The peak surface potential reportedly decreases from 0.4 V to -0.06 V in 84 minutes.[36] In contrast, the charged region on ultrathin glass retains its shape over time. Here, a gradual decrease in surface potential is observed without noticeable change in lateral spreading from its initial 2 μm dimension, indicating that charge dissipation occurs primarily through the bulk rather. This contrast in discharging behavior may be attributed to the difference in the composition and structure of the substrate. Unlike a thermally grown $SiO_2$ thin film, the presence of ions inside a stand-alone glass substrate can affect the surface charging and subsequent discharge behavior.

Further analysis of $\Delta V_{CPD}$ over time (Figure 2c) reveals an exponential charge decay behavior. The change in $\Delta V_{CPD}$ over time $t$ can be described by the exponential decay fit:

$$\Delta V_{CPD}(t) = \Delta V_\infty + (\Delta V_0 - \Delta V_\infty)e^{-t/\tau} \qquad (2)$$

where the initial contact potential difference $\Delta V_0 = 4.247 \pm 0.093$ V, the asymptotic baseline is marked by $\Delta V_\infty = 0.487 \pm 0.035$ V, and the decay time constant $\tau = 41.1 \pm 2$ min, with $R^2 = 0.9899$. After five decay time constants (~205 min), the calculated $\Delta V_{CPD}$ signal decays to 0.51 V which is within 1% of the asymptotic long-time baseline, and is in close agreement with the experimental $\Delta V_{CPD}$ of 0.448 V.

Under the ultra-pure $N_2$ conditions (Airgas research grade), where surface conduction via adsorbed water is strongly suppressed, the relatively longer time constant may be explained by a higher bulk resistivity of the stand-alone glass. The absence of lateral broadening of the charged area further suggests that the discharge occurs primarily through the bulk rather than along the surface. This time-dependent study establishes a base for the analysis of charge retention and dissipation behavior in standalone glass. Varying parameters such as glass composition, thickness, relative humidity, and ambient temperature, can provide a further quantitative insight into the factors affecting CE on glass at a nanoscale.

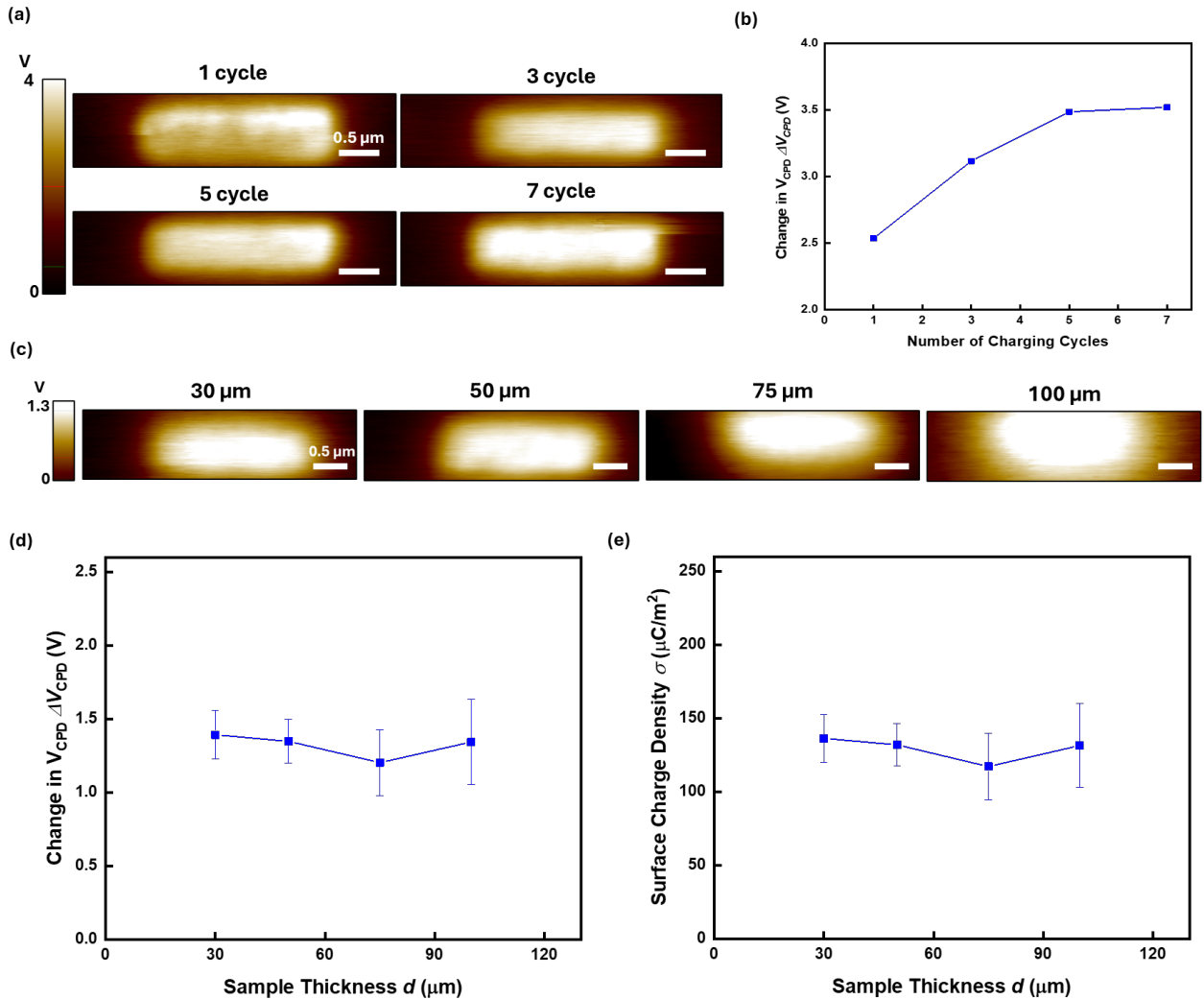

**Figure 3.** Effect of repeated charging cycle and glass thickness. (a) KPFM surface potential images of 30 μm glass after 1, 3, 5, and 7 contact cycles. (b) Change in $\Delta V_{\text{CPD}}$ with increasing number of charging cycles. (c) KPFM surface potential images for 30 μm, 50 μm, 75 μm, and 100 μm glass thicknesses. (d) Change in $\Delta V_{CPD}$ with increasing glass thickness. (e) Change in estimated surface charge density ($\sigma$) with increasing glass thickness.

To analyze the effect of repeated contact cycles on surface charging, the 30 μm sample was tested with 1, 3, 5, and 7 contact charging cycles before KPFM imaging. As shown in Figure 3a-b, $\Delta V_{CPD}$ increased with the number of cycles, rising from 2.54 V after a single cycle to 3.12 V after three cycles and 3.49 V after five cycles. Beyond this point, the response reaches a saturation point with seven cycles producing a comparable value of 3.52 V. Due to insufficient tip-sample contact, the distribution of charges after one contact cycle is not uniform. With increasing number of charging cycles, a more uniform charge distribution can be obtained.

Stand-alone ultrathin Corning glass samples with thicknesses of 30, 50, 75, and 100 μm were tested to analyze the effect of glass thickness on contact electrification (Figure S 3). As shown in Figure 3c-d, the measured $\Delta V_{CPD}$ remains constant across this thickness range, with values of 1.39 ± 0.17 V at 30 μm, 1.35 ± 0.15 V at 50 μm, 1.19 ± 0.23 V at 75 μm, and 1.34 ± 0.29 V at 100 μm. These results indicate that, within the ultrathin regime up to 100 μm, the proposed surface treatment enables consistent AFM-induced contact charging is largely insensitive to glass thickness.

The charge density at the sample surface ($\sigma$) can be approximated using a self-capacitance model of the rectangular sample surface charge. This self-capacitance dominates over the bottom electrode capacitance because the sample surface charge width $W$ and length $L$ are much smaller than all glass thicknesses studied in this report. As an approximation to the self-capacitance, we use the analytic formula for an equipotential thin circular disk of radius $R$:[45]

$$C_{disk}^{self} = 8\varepsilon_r \varepsilon_0 R \qquad (3)$$

The approximate value of $R$ is approximated by assuming the same circular area as the rectangular geometry, i.e., $\pi R^2 = LW$. The presence of air above the glass sample reduces the relative permittivity, which we approximate by the volumetric average of $\varepsilon_r \to \frac{1}{2}(\varepsilon_r^{glass}+1)$. The capacitance is then approximated as:

$$C^{self} \cong 4(\varepsilon_r^{glass} + 1)\varepsilon_0 \frac{\sqrt{LW}}{\sqrt{\pi}} \quad (4)$$

So, the conversion of $\Delta V_{CPD}$ voltage to charge density $\sigma$ becomes:

$$\sigma \cong \frac{C^{self}\Delta V_{CPD}}{LW} \cong \frac{4(\varepsilon_r^{glass}+1)\varepsilon_0}{\sqrt{\pi LW}} \Delta V_{CPD} \quad (5)$$

where $\varepsilon_r^{glass}$ is the relative permittivity of glass (~3.9 for SiO$_2$), $\varepsilon_0$ is the absolute permittivity of free space (8.854 × 10$^{-12}$ Fm$^{-1}$). As the sample thickness increases, it is found that $\sigma$ remains relatively constant over all thicknesses, with values of 136.26 ± 16.25 μC/m² at 30 μm, 131.93 ± 14.54 μC/m² at 50 μm, 117.15 ± 22.45 μC/m² at 75 μm, and 131.44 ± 28.41 μC/m² at 100 μm. The surface charge densities are comparable in magnitude to the previously reported ~70 μC/m² for the 200 nm SiO$_2$ thin film.[36] The surface charging response of various oxide, polymer, and semiconductor based substrates are compared in Table S 4.

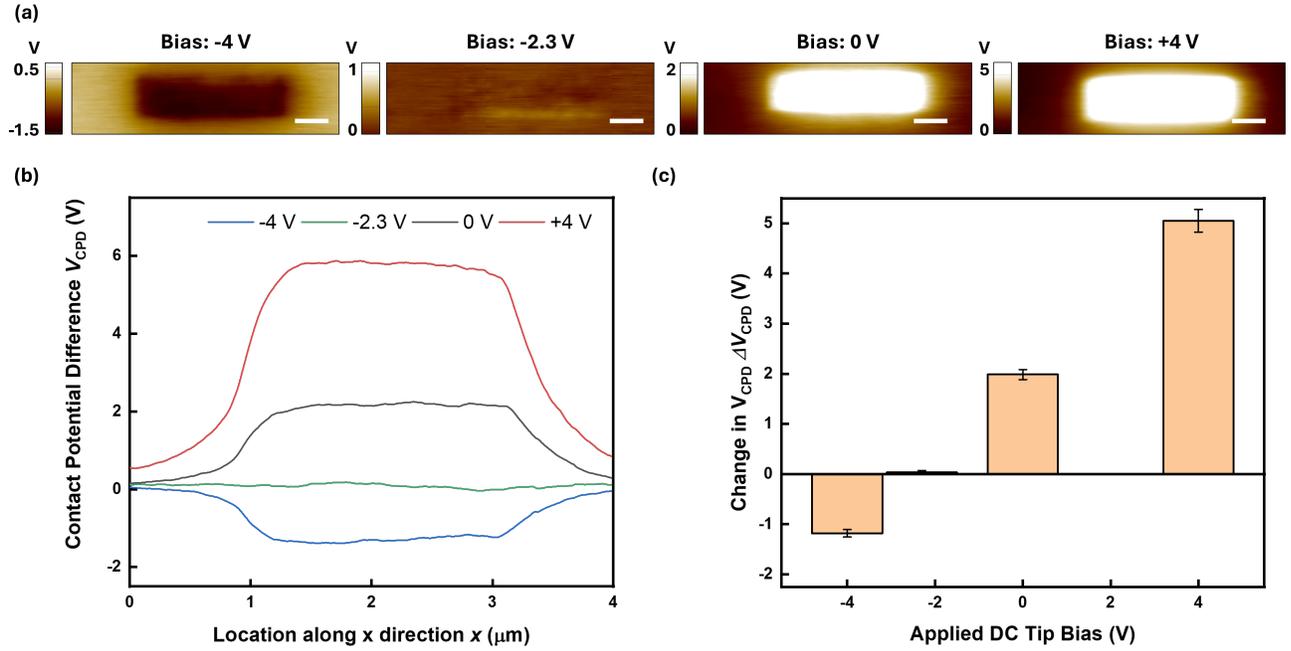

**Figure 4.** Effect of external electric field on CE with an applied DC tip bias of -4 V, -2.3 V, 0 V, +4 V during the contact charging scan. (a) KPFM surface potential images of 30 μm sample. (b) $V_{CPD}$ line profile along central line scan for 30 μm sample. (c) Change in $\Delta V_{CPD}$ for applied DC tip bias of -4 V, -2.3 V, 0 V, +4 V.

The effect of an external electric field on CE was also analyzed. The charging response was recorded for an applied tip bias of -4 V, -2.3 V, 0 V, and +4 V during one contact cycle. When a positive bias was applied, an enhancement of surface charging was observed, with a measured shift in $\Delta V_{CPD}$ close to +4 V. As shown in Figure 4, $\Delta V_{CPD}$ for the 30 μm sample increased from 1.98 ± 0.10 V for the 0 V bias case to 5.05 ± 0.23 V for +4 V bias. Minor discrepancies between the expected and the measured $\Delta V_{CPD}$ values can be attributed to imperfect tip-sample contact, and the natural decay of surface charges in the downtime between contact charging and subsequent KPFM imaging. For a negative DC tip bias case, $\Delta V_{CPD}$ decreased from 1.98 ± 0.10 V for 0 V bias to -1.18 ± 0.07 V for -4 V applied bias. The suppression and subsequent flipping of polarity of intrinsic charging can thus be observed with the applied negative bias. By applying a bias of -2.3 V, the intrinsic charging due to CE was completely nullified to a $\Delta V_{CPD}$ of 0.03 V. Thus, the charging of glass surfaces due to uncontrolled CE commonly observed in glass panel manufacturing, can be precisely controlled by an applied electric field to enhance, suppress, or invert the surface charges. In future studies, this analysis can be expanded further by systematically varying glass composition, thickness, and ambient conditions to study the combined effect of intrinsic CE and external bias on charge injection and stability in dielectric materials.

**CONCLUSION**

In summary, this work demonstrates the application of Kelvin probe force microscopy (KPFM) to systematically investigate contact electrification in stand-alone ultrathin glass with micrometer thickness. By combining the controlled AFM-based charge injection technique with a KOH based surface treatment, reproducible and stable nanoscale visualization of the charging and discharging behavior of the glass was achieved. The time-dependent measurements under ultra-pure $N_2$ conditions revealed a capacitor like decay behavior with a relatively long decay constant, which is different from the lateral charge dissipation behavior on thermally grown oxide in previous reports. Furthermore, by increasing the number of contact charging cycles, an increase in $\Delta V_{CPD}$ was observed up to five cycles after which a saturation in charging response was observed. A relatively constant $\Delta V_{CPD}$ was measured with increasing glass thicknesses from 30 µm to 100 µm. A self-capacitance model was developed to quantitatively estimate the thickness-dependent CE charge density. Finally, external electric fields were used to enhance, suppress, and invert the intrinsic CE response, offering a potential solution for active charge control during contact electrification.

## ASSOCIATED CONTENT

Histogram analysis of KPFM surface potential image of 30 μm glass sample after three contact charging cycles; time-lapse KPFM surface potential images of the same area on a 30 μm glass sample recorded over four hours after three contact charging cycles; KPFM surface potential images for 30 μm, 50 μm, 75 μm, and 100 μm glass thicknesses; surface charge behavior in oxide, polymer, and semiconductor substrates based on AFM induced nanoscale contact charging and subsequent KPFM measurements.

## AUTHOR INFORMATION

**Corresponding Author**

**Jun Liu** - Department of Mechanical and Aerospace Engineering and RENEW (Research and Education in Energy, Environment and Water) Institute, University at Buffalo, The State University of New York, Buffalo, New York 14260, United States

Email: jliu238@buffalo.edu

**Gabriel Agnello** – Science & Technology Division, Corning Incorporated, Corning, United States

Email: agnellogp@corning.com

**Authors**

**Aayush Nayyar** - Department of Mechanical and Aerospace Engineering and RENEW (Research and Education in Energy, Environment and Water) Institute, University at Buffalo, The State University of New York, Buffalo, New York 14260, United States

**Andrew C. Antony** – Science & Technology Division, Corning Incorporated, Corning, United States

**Dean Thelen –** Science & Technology Division, Corning Incorporated, Corning, United States

**Mayukh Nath –** Science & Technology Division, Corning Incorporated, Corning, United States

**Ruizhe Yang –** Department of Mechanical and Aerospace Engineering and RENEW (Research and Education in Energy, Environment and Water) Institute, University at Buffalo, The State University of New York, Buffalo, New York 14260, United States

† (Present Address) Pritzker School of Molecular Engineering, The University of Chicago, Chicago, Illinois 60637, United States

**Vashin Gautham -** Department of Mechanical and Aerospace Engineering and RENEW (Research and Education in Energy, Environment and Water) Institute, University at Buffalo, The State University of New York, Buffalo, New York 14260, United States

**Sagnik Das -** Department of Chemical and Biological Engineering, University at Buffalo, The State University of New York, Buffalo, New York 14260, United States

**Haiqing Lin -** Department of Chemical and Biological Engineering, University at Buffalo, The State University of New York, Buffalo, New York 14260, United States

**Author Contributions**

J.L. conceived the idea of the work. A.N. and J.L. designed and conducted the experiments. A.N., G.A., A.A., R.Y., V.G., and J.L. designed the surface treatment procedure. A.N., D.T., M.N., and J.L. worked on time dependent decay of charges and self-capacitance model, A.N., S.D., and H.L. carried out the contact angle measurements. All authors contributed to the discussion and editing of the manuscript.

**Funding Sources**

**Notes**

The authors declare no competing financial interest.

## ACKNOWLEDGEMENTS

This study was supported by Corning Incorporated and UB's Center of Excellence in Materials Informatics (CMI) Faculty-Industry Applied Research Opportunities Grant (FIAR).